\documentclass[superscriptaddress]{article}
\usepackage{authblk} 
\usepackage[utf8]{inputenc}
\usepackage{amsmath, amsthm, amssymb, amsfonts}
\usepackage{mathrsfs}
\usepackage{dsfont}
\usepackage{amsthm}
\usepackage{csquotes}
\usepackage[hyperfootnotes=false]{hyperref}

\begin{document}
	
\title{Comment on `The aestivation hypothesis for resolving Fermi’s paradox'}
\date{\today}
\renewcommand\Affilfont{\itshape\small}
\author[1]{Charles H. Bennett}
\author[2]{Robin Hanson\footnote{Email: rhanson@gmu.edu}}
\author[3]{C.\ Jess Riedel\footnote{Email: jessriedel@gmail.com}}
\affil[1]{IBM Watson Research Center, Yorktown Heights, NY, USA}
\affil[2]{Department of Economics, George Mason University, Fairfax, VA, USA}
\affil[3]{Perimeter Institute for Theoretical Physics, Waterloo, Ontario, Canada}

\maketitle

\begin{abstract}
In their article, ‘That is not dead which can eternal lie: the aestivation hypothesis for resolving Fermi’s paradox’, Sandberg et al.\ try to explain the Fermi paradox (we see no aliens) by claiming that Landauer’s principle implies that a civilization can in principle perform far more (${\sim} 10^{30}$ times more) irreversible logical operations (e.g., error-correcting bit erasures) if it conserves its resources until the distant future when the cosmic background temperature is very low. So perhaps aliens are out there, but quietly waiting. Sandberg et al.\ implicitly assume, however, that computer-generated entropy can only be disposed of by transferring it to the cosmological background. In fact, while this assumption may apply in the distant future, our universe today contains vast reservoirs and other physical systems in non-maximal entropy states, and computer-generated entropy can be transferred to them at the adiabatic conversion rate of one bit of negentropy to erase one bit of error. This can be done at any time, and is not improved by waiting for a low cosmic background temperature. Thus aliens need not wait to be active. As Sandberg et al.\ do not provide a concrete model of the effect they assert, we construct one and show where their informal argument goes wrong.
\end{abstract}

\section{Introduction}
In this note we critique the thermodynamic claims made by Sandberg et al.\ in their article  `That is not dead which can eternal lie: the aestivation hypothesis for resolving Fermi’s paradox' \cite{sandberg2017that}.  Our main point is related to these quotes: 
\begin{displayquote}
	\textit{The argument is that the thermodynamics of computation make the cost of a certain amount of computation proportional to the temperature...As the universe cools down, one Joule of energy is worth proportionally more. This can be a substantial ($10^{30}$) gain...at least $E \ge kT \ln(2)$J need to be dissipated for an irreversible change of one bit of information...It should be noted that the thermodynamic cost can be paid by using other things than energy. An ordered reservoir of spin or indeed any other conserved quantity can function as payment...However, making such a negentropy reservoir presumably requires physical work unless it can be found naturally untapped...irreversible operations must occur when new memory is created and in order to do error correction...A comparison of current computational resources to late era computational resources hence suggest a potential multiplier of $10^{30}$!...For the purposes of this paper we will separate the resources into energy resources that can power computations and matter resources that can be used to store information, process it...}
\end{displayquote}
We adopt their terminology, using ``civilization'' to refer to an arbitrarily technologically powerful agent in the universe, ``reservoir'' to refer to a bounded thermodynamic system (e.g., a battery) that can be manipulated by a civilization, and ``bath'' to refer to an effectively infinite thermalized system whose temperature is exogenously determined (e.g., the cosmic microwave background) \cite{terminology}.  We understand their argument to be as follows [our words]: 
\begin{displayquote}
	The fundamental thermodynamic resource of computation is negentropy, but for practical purposes this can be measured in energy (because accessible ordered reservoirs of conserved quantities besides energy appear to be a small fraction of all naturally occurring negentropy).  A future civilization seeking to perform very extensive computations within thermodynamic constraints may utilize reversible computing to minimize the amount of thermodynamic expenditure per unit of computation, but it is unavoidable that they must correct a residual rate of physical error.   Correcting these errors ultimately requires bit erasures, whereby the entropy is ejected into an external system with an energy cost given by Landauer's principle.  Efficiency is maximized when the system is at the temperature of the coldest available natural bath, which in the future will be that of the cosmological background.  Although civilizations may collect energy at early times, they reap a much larger computational harvest if they store that energy and spend it at later times when the cosmic background is much lower.
\end{displayquote}
We are grateful to A. Sandberg for confirming that this is a fair summary of the argument~\cite{sandberg2018private}.

This informal argument conflicts with the intuitive notion that the fundamental spendable resource for irreversible operations like bit erasure is negentropy and the conversion rate is very simple: one bit of negentropy erases one bit of error, regardless of the temperature of external baths \cite{bennett1982thermodynamics}.  As we will explain, it appears Sandberg et al.\ err by implicitly assuming that the entropy generated by bit erasure cannot be transfered into \emph{any} bounded reservoir and instead \emph{must} be ejected into a particular unbounded bath, in this case the cosmic microwave background (CMB).  It is possible to build a toy model that embodies this assumption, leading to thermodynamic incentives for agents to wait (``aestivate''~\cite{spelling}) until a future colder period to perform irreversible operations like bit erasure, but that model does not approximate our universe.  More specifically there is no incentive to delay irreversible operations until 
\begin{enumerate}
	\item the civilization has taken control of all accessible matter in the universe in the sense that it can inhibit all non-adiabatic physical interactions from occurring, both within the matter and between the matter and the CMB; and 
	\item that accessible matter has been fully exploited in the sense that it has internally thermalized, i.e., non-gravitational heat death, perhaps after eons of computations.
\end{enumerate}
Furthermore, the aestivation incentive in such an expanding quasi-thermalized universe is not at all specific to computation, but rather applies to any activity that requires negentropy.

In Appendix~\ref{other-criticism} we make some other, less important comments about discussion in Ref.~\cite{sandberg2017that}.

\section{Our model}

To see how the informal reasoning of Sandberg et al.\ breaks down, we need a concrete model of their purported effect.  Unfortunately, Ref.~\cite{sandberg2017that} does not provide enough detail to unambiguously specify a model, so we do our best to translate their informal description into the non-cosmological model below.  (We are grateful to the authors in assisting us through email correspondence, but the following exposition should not be interpreted as being endorsed by them.)

We take the universe to consist of these thermodynamic components: 
\begin{enumerate}
	\item 
	A memory tape of $N$ bits.  These bits are initially all zero, but they are sequentially randomized as the civilization performs computations (outside the model) that generate occasional errors, which are swapped onto the memory tape.  Once the memory tape is full of randomized bits, they must be erased to allow additional computations.
	\item
	An unusual reservoir that has the capacity to provide a large finite amount of energy in the form of useful work \emph{but that cannot absorb appreciable entropy}.  This could be a system with a tiny number of states with extremely large energy differences that is internally thermalized to a temperature much higher than the temperatures of all other systems.
	\item
	A thermal bath of effectively infinite size, following an exogenously set temperature schedule $T(t)$, with which the civilization can exchange an arbitrary amount of heat.
	\item
	A bit-erasing machine that also has a negligible number of internal states.  If work is supplied to the bit eraser, it pumps entropy from the memory tape to the bath.  This can be formalized in at least two ways:
	\begin{enumerate}
		\item
		taking the bit eraser to be a Szilard engine augmented so that it can perform a reversible swap operation between a memory bit and the binary location (left or right) of the gas particle in the engine \cite{bennett1982thermodynamics}; or  
		\item
		assuming the different memory configurations actually have very small energy differences, and taking the bit erasers to be a reversible Carnot engine that ``cools'' the memory tape to a sufficiently cold temperature that all bits are zero with high probability.
	\end{enumerate}
\end{enumerate}
In particular, this is a non-relativistic model that can be analyzed with the standard techniques of classical thermodynamics, the setting in which Landauer's principle is traditionally derived.  Following Sandberg et al., we have assumed the key thermodynamic features of an expanding spacetime are fully captured by the thermal-bath temperature schedule $T(t)$ as a model for the CMB, and so are ignoring potentially important features like the increasing volume, decreasing pressure, and decreasing particle density of the real CMB as it evolves into the future.  In particular, we emphasize the possibility that a finite effective bath size (e.g., from the finite number of CMB photon modes\footnote{Note that if the photon mass is non-zero, this will become relevant at the extremely cold temperatures, $k_B\cdot(10^{-31}\,\mathrm{K}) \sim 10^{-35}\,\mathrm{eV/c^2}$, necessary to obtain the computational enhancements Sandberg et al.\ assert.  The experimental upper bound on the photon mass is $10^{-18}\,\mathrm{eV/c^2}$ \cite{pdg}.} accessible from a finite volume) may invalidate conclusions drawn from this crude model, mooting both the arguments of Sandberg et al.\ and our rebuttal.  

The unusual reservoir (\#2) is a crucial assumption whose applicability to the actual universe we will later dispute, but we provisionally accept it so that we can now exhibit Sandberg et al.'s conclusions within our model.  In doing so, we will derive Landauer's principle from the more fundamental thermodynamic laws of conservation of energy and non-decrease of total entropy.  

Suppose a large number $N$ of bits of the memory tape are random and need to be erased, requiring the removal of an amount of entropy $\Delta S = N \ln 2$. Then, by the second law, the entropy of the bath must increase by at least this amount because, by assumption, neither the reservoir nor the bit eraser have an appreciable number of internal states.  Since the bath is at maximum entropy given its internal energy, an amount of heat energy $\Delta Q = T \Delta S$ must be added to the bath -- by the definition of temperature -- in order for its entropy to increase by the necessary amount.  And by conservation of energy this must be supplied as work $W = \Delta Q = T N \ln 2$ from the reservoir.  

The conclusions of Sandberg et al.\ then follow: if $T(t)$ is decreasing with time, the civilization performs more total bit erasures before the work reservoir is exhausted (and so, by assumption, more total computations) if it waits until a later time when $T$ is lower. 

\section{Critique of assumption}

The key feature necessary for the above conclusions is that the reservoir cannot accept an appreciable amount of entropy despite being able to do prodigious amounts of work. This appears in Ref.~\cite{sandberg2017that} as an assumption that all of the reservoir's internal energy is available to do useful work, a conflation of the energy with the \emph{free} energy.  As we will now show, the incentive to aestivate disappears when we drop the unreasonable assumption that exclusively reservoirs of this sort are accessible to the civilization.

The real universe is full of subsystems, besides the CMB, that are out of equilibrium with each other and therefore can accept additional entropy.  The most general possible reservoir has an entropy $S$ and an internal energy $U$ that is associated with some maximum entropy $S_{\mathrm{max}}(U)$ determined by physical properties of the reservoir.  These physical properties include things like the species of particles it is constructed from, the charges of any conserved quantities, and the maximum volume the reservoir can occupy while maintaining its physical integrity.  If $S=S_{\mathrm{max}}$ then the reservoir has thermalized.  Otherwise, if $S$ is less than $S_\mathrm{max}$, then there generally exists a reversible transformation that moves the entropy in the memory to the reservoir.\footnote{It's possible to construct counterexample for which $S + N\ln 2 < S_\mathrm{max}$ but there do not exist reversible physical transformations, formalized as Hamiltonian flow in the joint memory-reservoir phase space, that move all entropy from the memory to the reservoir.  However, this can only be done by appealing to constraints that do not follow from the first or second laws of thermodynamics.  This is not relevant in the present context because we are rebutting the thermodynamic arguments of Sandberg et al., and because matter content of the actual universe clearly has the ability to absorb huge amounts of entropy without ejecting it into the CMB.}  Since all memory states have the same energy, or nearly so, the total energy of the reservoir does not change.

To illustrate this, assume for simplicity that the reservoir is composed of multiple discrete parts that each have a well-defined temperature. Consider first just two parts at different temperatures $T_1$ and $T_2$. The parts can be connected by a Carnot engine that generates work powered by allowing heat to flow from the hotter part to the colder one.  So long as $T_1 \neq T_2$, the Carnot engine can power the bit eraser, reducing the temperature differential, to reversibly pump entropy from the memory tape into the colder reservoir part.  (Ejecting the entropy into a locally thermalized part of the reservoir in accordance with Landauer's principle is no different than ejecting it into the CMB bath.)  This process can continue until either the memory tape is blank or the temperature difference is exhausted ($T_1 = T_2$), so that no further work can be extracted by the Carnot engine.  The maximum number of bits erased is set by the difference between the total initial entropy of the two parts and their total final entropy when they have thermalized at the joint temperature set by conservation of energy.  Since this is a reversible process, this bit-erasing capacity is ideal.

As more random bits -- generated by external computations -- are swapped into the memory, one can keep erasing bits in the memory until \emph{all} parts of the reservoir have equilibrated to the same temperature, i.e., the reservoir is at maximal entropy for its energy.  The maximum number of erasures that we can make is given by $\Delta S/ \ln 2 = (S_{\mathrm{max}} - S)/ \ln 2$, i.e., the initial negentropy of the reservoir measured in bits.

Importantly, the erasures are reversible and we have not made use of any interaction with the CMB bath, so the erasures can be made at any time without regard for the changing bath temperature. It is only if the civilization irreversibly pushes entropy into the uncontrolled CMB bath at an inopportunely high temperature that it ``eats the seed corn'', sacrificing a possible aestivation bonus in accordance with the previous section.

Thus we come to our first conclusion: a civilization can freely erase bits without forgoing larger future rewards up until the point when all accessible bounded resources are jointly thermalized.  At that time, the contents of the universe would appear to be in equilibrium (heat death), much like the early universe before structure formation.  This is very different from the present universe, so there is currently no incentive for aliens to aestivate.  In fact, as show in the next section, quite the opposite.

\section{Reservoir-Bath coupling}

Now let us assume that the reservoir has been exhausted (internally thermalized) as described in the previous section. (This situation differs from our initial toy model in that the reservoir, on its own, can do no useful work.) We suppose civilization desires to power further bit erasures by exploiting the temperature \emph{difference} between the bath and the reservoir that is induced by the changing bath temperature $T(t)$, i.e., powering the bit eraser with heat flow between the reservoir and the bath. Within this model, it is necessary for the civilization to wait for the bath temperature to fall as far below the reservoir temperature as possible to maximize the number of erasures performed.  

To see this explicitly, let us assume that the finite reservoir has a constant heat capacity $C$ and initial temperature $T_R$.  The infinite CMB bath temperature $T(t)$ falls from initial temperature $T_i$ to final temperature $T_f \ll T_i$ at a rate that is slow compared to the speed at which the bit eraser can exhaust the reservoir.
We show in Appendix~\ref{reservoir-capacity} that, for quasi-static bath temperature $T$ and reservoir temperature $T_R>T$, a maximum of 
\begin{align}
\label{reservoir-erasures}
N(T,T_R) = \frac{C}{k \ln 2}\left(\frac{T_R}{T} - 1 - \ln \frac{T_R}{T}\right)
\end{align}
bits can be erased before the reservoir equilibrates to the bath.   Now consider two strategies as the bath temperature falls from $T_i$ to $T_f$: the ``greedy'' strategy continuously exploits any temperature differential between the bath and the reservoir to perform erasures, while the ``patient'' strategy aestivates until $T = T_f$, running the bit eraser only once.  

In the first round of the greedy strategy, the civilization immediately performs $N(T_i,T_R)$ erasures, thereby lowering the reservoir to the same temperature $T_i$ as the bath.  Then after a while the bath temperature falls a small amount,  $T = T_i - \Delta T$, and the civilization performs a round of erasures which, for small $\Delta T$, yields
\begin{align}
N(T-\Delta T,T) \approx \frac{C}{k \ln 2}\left[\frac{1}{2} \frac{\Delta T^2}{T^2} + O\left(\frac{\Delta T^3}{T^3}\right)\right] \propto \left(\frac{\Delta T}{T}\right)^2 \le \left(\frac{\Delta T}{T_f}\right)^2
\end{align}
bits erased.
The civilization continues to perform subsequent rounds of erasures each time the temperature falls $\Delta T$.  The number of rounds that are performed before $T=T_f$ increases linearly with the inverse step size $\Delta T^{-1}$, but the number of erasures per round is proportional to $\Delta T^{2}$, so the total number of erasures performed in these rounds vanishes in the continuous (maximally greedy) limit $\Delta T \to 0$.  Therefore, the greedy strategy yields only the $N(T_i,T_R)$ erasures from the initial round.

On the other hand, for the patient strategy, only a single round of erasures are performed after the bath has reached it's final temperature, yielding $N(T_f,T_R)$ erasures total, an improvement on the greedy strategy.  Assuming a very cold bath, $T = T_f \ll T_R$, the number of erasures scales proportional to the inverse final temperature of the bath:
\begin{align}
  N(T_f,T_R) \overset{T/T_R \to 0}{\approx} \frac {C\, T_R}{k \, T_f \ln 2} \propto \frac{1}{T_f}.
\end{align}
In agreement with the claims of Sandberg et al., we see that -- once the reservoir has been fully thermalized and only the reservoir-bath differential remains to be exploited -- the civilization is incentivized to perform all computations and erasures in the distant future when the bath temperature $T$ has stabilized to its minimum temperature $T_f$ (assuming the reservoir is sufficiently insulated).

But here again is where the implicit assumptions by Sandberg et al.\ (at least as we have interpreted them) can be disputed.  The basic idea is that the civilization could power additional erasures -- without the need for aestivation -- by simply corralling part of the infinite bath and treating it as an additional finite reservoir.  Assuming only access to a small amount of inert insulating matter (which would have no thermodynamic value in a non-expanding universe) the civilization can build a large container that is empty except for the blackbody radiation it contains (which is initially equivalent to CMB radiation by virtue of the temperature of the walls).  Now the civilization has two reservoirs, each internally thermalized at different temperatures.  A reversible Carnot engine exploiting their temperature difference can power the bit eraser, pumping entropy from the memory into the cooler reservoir, until the two reservoirs are at the same temperature, as described in the previous section.  This process -- converting some of the bath into a new reservoir and then mining the temperature difference between reservoirs until they equilibrate -- can be repeated for as long as there is any matter in the universe accessible to the civilization.\footnote{Note that the proton lifetime is constrained to be greater than ${\sim}10^{29}$ years\cite{pdg}, much larger than the ${\sim}10^{12}$-year timescale on which the effective temperature of the CMB bath reaches its fixed point on account of CMB photons redshifting below the de-Sitter temperature $T_f \equiv T_{\mathrm{dS}} \approx 2.7\cdot 10^{-30}$ K \cite{ziblin2007evolution}.}  Just as before, the civilization can use this process to freely make bit erasures without forgoing an aestivation bonus.  The key idea is that by pushing entropy into the corralled photons in the new reservoir, rather than the uncontrolled photons in the CMB, it's always possible to reverse the transformation.  Only once the civilization has commandeered all accessible matter in the universe will they face the incentive to aestivate since they now \emph{must} push bit-erasure entropy into the uncontrolled bath. 

Thus, even beyond the normal incentive to acquire sources of negentropy (out-of-equilibrium reservoirs), the falling temperature schedule of the CMB is an \emph{additional} incentive for the civilization to take control of as much of the universe as possible -- even the parts that are thermalized and inert!

\section{Final comments}

We have concluded that a civilizations capable of reversible computing has no incentive to aestivate until after it has taken control of, and fully exploited, all accessible matter in the universe.  

None of this is specific to computation.  The incentive to wait, or lack thereof, applies just as well to a civilization whose terminal desires involve expending work to move matter into particular configurations (e.g., galactic-scale art projects) whose limiting cost is residual frictions, analogous to residual computational faults necessitating error correction.

We have not addressed many other potential issues that would arise in waiting until the far future of the actual universe, such as the decreasing speed with which thermalization with the CMB bath can happen as the photon density gets lower, the ultimate limits of insulation, and complications from the finite size of atoms.  It's unclear whether the aestivation incentive will persist, even in the specific scenario discussed in the previous section, given a more realistic cosmological model.

It has not escaped our notice that dark matter, though its nature remains a mystery to us, may have a purpose for a powerful civilization able to produce it: to temporarily sequester most of the universe's mass in a form whose dynamics conserves not only energy and angular momentum, but to a good approximation entropy, thereby saving it until it can be most expeditiously converted back to ordinary matter to power computation or other projects.  Of course, one might hope that such a civilization would preserve a few of the stars, supernovae, etc.\ as a grand public art project, beautiful and/or controversial, but consuming only a small fraction of all resources.

\appendix

\section{Other criticism}
\label{other-criticism}

Here we make additional and less important comments.  They are independent and so are not necessary to understand our main criticism above.

\subsection{Thermodynamic costs of computation beyond erasure}

Consider these quotes:
\begin{displayquote}
	\textit{If advanced civilizations do all their computations as reversible computations, then it would seem unnecessary to gather energy resources (material resources may still be needed to process and store the information). However, irreversible operations must occur when new memory is created and in order to do error correction...the actual correction is an irreversible operation...Error rates are suppressed by lower temperature and larger/heavier storage. Errors in bit storage occur due to classical thermal noise and quantum tunneling.}
\end{displayquote}
Our most physically realistic models of computation, Brownian computers, have another big entropy cost of computation: ``friction'' due to driving motion ``forward'' at a finite rate \cite{bennett1982thermodynamics,bennett1973logical}. This entropy cost per gate operation goes inversely as the time taken per gate operation. If errors happen at a constant physical rate, then trading these costs sets an entropy-cost minimizing time period per gate operation. If stored negentropy were the limiting resource, and not for example computer hardware, then this would set an optimal rate for using negentropy.

Sandberg et al.\ instead treat error correction as the only entropy cost, and thus say that the min entropy compute strategy is to wait until the universal background temperature reaches a low level.

\subsection{Reversibility of cooling}

This sentence in Ref.~\cite{sandberg2017that} is incorrect, possibly for similar reasons as we discuss above:
\begin{displayquote}
	\textit{While it is possible for a civilization to cool down parts of itself to any low temperature, the act of cooling is itself dissipative since it requires doing work against a hot environment.}
\end{displayquote}
Cooling does not need to be dissipative.  That is, cooling a system requires negentropy but it does not necessarily destroy it; the negentropy used can be recovered if the system is allowed to warm up again.  For instance, given a charged battery and two thermal reservoirs at the same temperature, negentropy can be extracted from the battery (in the form of an applied electromotive force, discharging the battery) and transfered to the reservoirs using a reversed Carnot cycle that pumps heat from the one to the other (resulting in a net temperature difference between them).  This process is adiabatic and hence reversible.

Thus the fact that error rates can rise with temperature is a reason to run a computer at a low temperature, but not necessarily a reason to wait for low universal background temperatures. 

Likewise:
\begin{displayquote}
	\textit{The most efficient cooling merely consists of linking the computation to the coldest heat-bath naturally available.}
\end{displayquote}
Allowing heat to directly flow from something warm (the computational machinery) freely to something cool (the bath) unnecessarily increases entropy, and so is not the most efficient method.

\section{Total work extractable from a reservoir-bath differential}
\label{reservoir-capacity}

Here we calculate how much total useful work can be extracted from a reversible engine (e.g., Carnot) operating between an infinite thermal bath at temperature $T$ and a finite thermalized reservoir at initial temperature $T_0>T$ if the reservoir is assumed to have constant heat capacity $C$.  It is known that the infinitesimal work dW generated by a Carnot engine is related to the heat leaving a reservoir $dQ_R$ and the heat entering the bath $dQ_B$ by the relations
\begin{align}
dQ_R &= dQ_B + dW \qquad &\mathrm{(conservation\, of\, energy)} \\
\frac{dQ_R}{dQ_B} &= \frac{T_R}{T} \qquad &\mathrm{(reversible\, heat\, engine)}\, \\
C &= \frac{dQ_R}{dT_R} \qquad &\mathrm{(constant\, heat\, capacity)}
\end{align}
where $T_R$ is the instantaneous temperature of the reservoir.  The total work is obtained by integrating dW from the initial condition $T_R = T_0$ to the asymptotic final state $T_R = T$:
\begin{align}\begin{split}
W &= \int dW = \int (dQ_R - dQ_B) = \int dQ_R (1-T/T_R) \\
&= C \int^{T_0}_T dT_R (1-T/T_R) = C T (\beta - 1 - \ln \beta)
\end{split}\end{align}
where $\beta \equiv T_0/T$.  This work can be used to power the bit eraser at the Landauer limit, yielding 
\begin{align}
N = \frac{W}{k \, T \ln 2} = \frac{C}{k \ln 2}(\beta - 1 - \ln \beta)
\end{align}
erasures.

\end{document}